\documentstyle[prd,aps]{revtex}

\newcommand{\bea}{\begin{eqnarray}}
\newcommand{\eea}{\end{eqnarray}}

\begin{document}

\draft

\title{Gravitational Wave Spectrums from Pole-like Inflations based on 
       Generalized Gravity Theories}
\author{Jai-chan Hwang}
\address{Department of Astronomy and Atmospheric Sciences,
         Kyungpook National University, Taegu, Korea}
\date{\today}
\maketitle

\begin{abstract}

We present a general and unified formulation which can handle the classical 
evolution and quantum generation processes of the cosmological gravitational 
wave in a broad class of generalized gravity theories.  
Applications are made in several inflation models based on the scalar-tensor 
theory, the induced gravity, and the low energy effective action of string 
theories.
The gravitational wave power spectrums based on the vacuum expectation value 
of the quantized fluctuating metric during the pole-like inflation stages 
are derived in analytic forms.
Assuming that the gravity theory transits to Einstein one while the relevant 
scales remain in the superhorizon scale, we derive the consequent power 
spectrums and the directional fluctuations of the relic radiation produced 
by the gravitational wave.
The spectrums seeded by the vacuum fluctuations in the pole-like inflation 
models based on the generalized gravity show a distinguished common feature 
($n_T \simeq 3$ spectrum) which differs from the scale invariant one 
($n_T \simeq 0$ spectrum) generated in an exponential inflation in 
Einstein gravity which is supported by observations.

\end{abstract}



\section{Introduction}

Pioneering studies of the cosmological gravitational wave in Eintein 
gravity are made in \cite{Lifshitz,Sachs-Wolfe,Grishchuk,Ford-Parker}.
The inflation generated gravitational wave spectrums are widely studied
in the context of Einstein's gravity, 
\cite{pre-COBE,Abbott-Wise,Starobinsky,Bogoliubov}.
Recent detection of the large angular scale fluctuations in the
cosmic microwave background radiation spurred renewed interests
in the potential significance of the inflation generated cosmological
gravitational wave, \cite{post-COBE,Allen}.
{}For a historical review, see \cite{Allen}.

In a series of work we have been investigating the large scale 
structure formation processes in the context of early universe models
based on a broad class of generalized gravity theories. 
We have presented the classical evolution \cite{GGT-H} and the quantum 
generation \cite{GGT-QFT} processes for scalar type perturbation which 
later evolves to the spatial density fluctuation in the large scale.
Applications to scenarios with a few inflation models are
made in \cite{Kin}.
Besides the scalar type perturbation there exist two other types:
the vector and tensor types.
Due to the high symmetry of the background cosmological model 
(the spatial homogeneity and isotropy) these three types of perturbation
decouple from each other and evolve independently.
The transverse vector type perturbation corresponds to the rotational mode
which simply decays in an expanding medium due to the angular momentum 
conservation.
The transverse-tracefree tensor type perturbation corresponds to 
the gravitational wave mode which is preserved in the superhorizon scale 
and is redshifted away on scales inside horizon.
The equation describing the gravitational wave in the generalized gravity
has a similar structure compared with the spatial curvature fluctuation
of the scalar type perturbation in certain gauges; the equations and
comparisons are made in \cite{GGT-H}.
Previous studies of the cosmological gravitational wave in the context of the
generalized gravity can be found in 
\cite{GW-GGT,GGT-1990,PRW,MFB,pre-big-bang-GW,GGT-H}.

In this paper we will present the general and unified formulation 
(both classical and quantum) for handling the cosmological gravitational
wave in generalized gravity, and will apply the formulation to several 
inflation models based on generalized gravity theories.
This paper can be considered as a gravitational wave counterpart
of \cite{Kin} which was concerned with the scalar type perturbation
in generalized gravity.
Since the gravitational field equation and the cosmological background 
equation will overlap we recommend to read this paper together with \cite{Kin}.

In Sec. \ref{sec:Classical} we present the classical action for the
gravitational wave valid in a broad class of generalized gravity theories.
We present the general asymptotic solutions and an exact solution valid 
under a condition which is, in fact, very general so that it
includes most of the prototype inflation models known in Einstein gravity 
and generalized gravity.
In Sec. \ref{sec:Quantum} we present a quantum formulation and derive the 
normalization condition for the vacuum expectation value. 
In Sec. \ref{sec:Applications} we apply the formulation in 
Sec. \ref{sec:Quantum} to several expansion stages realized in generalized 
gravity theories. 
We derive the power spectrums based on the vacuum expectation.
In Sec. \ref{sec:Spectrums} we derive the consequent classical power spectrums
and the observational effects on the temperature anisotropy of the
cosmic microwave background radiation.
We use the conserved character of the growing mode of the gravitational wave 
in the large scale limit which applies independently of changes in the 
background equation of state and the gravity theories.
In Sec. \ref{sec:Discussions} comparisons are made with the power spectrums
derived in the scalar type perturbation, and a brief discussion is presented.

We set $c \equiv 1$.

\section{Classical evolution}
                               \label{sec:Classical}

We consider gravity theories included in the following action which we call 
a generalized $f(\phi,R)$ gravity \cite{GGT-1990}
\bea
   S = \int d^4 x \sqrt{-g} \left[ {1 \over 2} f (\phi, R)
       - {1\over 2} \omega (\phi) \phi^{;a} \phi_{,a} - V(\phi) \right].
   \label{GGT-action}
\eea
The various names of generalized gravity theories which are subsets
of this action are summarized in \cite{GGT-H}.
We consider a homogeneous and isotropic (flat) cosmological model
with the tensor type perturbation
\bea
   d s^2 = - d t^2 + a^2 \left( \delta_{\alpha\beta} 
       + 2 C^{(t)}_{\alpha\beta} \right) d x^\alpha d x^\beta,
   \label{metric-general}
\eea
where $C^{(t)}_{\alpha\beta}({\bf x}, t)$ is a transverse-tracefree 
($C^{(t)\alpha}_{\;\;\;\;\; \alpha} = 0 = C^{(t)|\beta}_{\alpha\beta}$)
tensor type perturbation corresponding to the gravitational wave;
the indices are based on $\delta_{\alpha\beta}$ as the metric.

The gravitational field equation and the equation of motion
for the general covariant case and for the background metric
are presented in Eqs. (2,5) of \cite{Kin}.
The second order perturbed action for the gravitational wave part of 
Eq. (\ref{GGT-action}) becomes 
\bea
   \delta^2 S_g = \int {1 \over 2} a^3 F
       \left( \dot C^{(t)\alpha}_{\;\;\;\;\beta} 
       \dot C^{(t)\beta}_{\;\;\;\;\alpha}
       - {1 \over a^2} C^{(t)\alpha}_{\;\;\;\;\beta,\gamma}
       C^{(t)\beta|\gamma}_{\;\;\;\;\alpha} \right) dt d^3 x,
   \label{GW-action}
\eea
where $F \equiv \partial f/ \partial R$ and an overdot indicates 
a time derivative based on $t$.
[Derivation of Eq. (\ref{GW-action}) is the following:
The action expanded to second order in $C_{\alpha\beta}^{(t)}$
is presented in Eq. (18.6) of \cite{MFB} in the case of Einstein gravity; 
see also \cite{Ford-Parker,Allen}.
Corresponding action in the case of the generalized $f(\phi,R)$
gravity can be derived by applying the conformal transformation
properties presented in \cite{GGT-CT}.
The result is the one in Eq. (\ref{GW-action}).]
The equation of motion is 
\bea
   \ddot C^{(t)}_{\alpha\beta}
       + \left( 3 H + {\dot F \over F} \right) \dot C^{(t)}_{\alpha\beta}
       - {1 \over a^2} \nabla^2 C^{(t)}_{\alpha\beta} = 0,
   \label{GW-eq-C}
\eea
where $H \equiv \dot a/a$; for an equation considering the general curvature
term in the background, see Eq. (102) in \cite{PRW}.

Equation (\ref{GW-eq-C}) can be written as
\bea
   & & v_g^{\prime\prime} 
       - \left( {z_g^{\prime\prime} \over z_g} + \nabla^2 \right) v_g = 0,
   \nonumber \\
   & & v_g ({\bf x}, t) \equiv a \sqrt{F} C_{\alpha\beta}^{(t)} ({\bf x}, t), 
       \quad z_g \equiv a \sqrt{F},
   \label{v-eq}
\eea
where a prime indicates the time derivative based on the conformal time
$\eta$ ($dt \equiv a d \eta$).
The large and small scale asymptotic solutions can be derived, respectively, as
\bea
   & & C_{\alpha\beta}^{(t)} ({\bf x}, t)
       = C_{\alpha\beta} ({\bf x}) - D_{\alpha\beta} ({\bf x}) 
       \int^t_0 {1\over a^3 F} dt,
   \label{GW-LS-sol} \\
   & & C_{\alpha\beta}^{(t)} ({\bf k}, \eta)
       = { 1 \over a \sqrt{F} } 
       \Big[ c_{1 \alpha\beta} ({\bf k}) e^{ik\eta} 
       + c_{2 \alpha\beta} ({\bf k}) e^{-ik\eta} \Big]. 
   \label{GW-SS-sol}
\eea
Notice that in the large scale limit (larger than the visual horizon)
the growing mode is conserved independently of the general changes 
in the background equation of state [i.e., for general $V(\phi)$] 
and of the changes in the gravity theories 
[i.e., for general $f(\phi,R)$ and $\omega(\phi)$].

{}For $z_g^{\prime\prime}/z_g = n_g/\eta^2$ with $n_g = {\rm constant}$, 
Eq. (\ref{GW-eq-C}) has an exact solution
\bea
   & & C_{\alpha\beta}^{(t)} ({\bf k}, \eta) 
       = {\sqrt{|\eta|} \over a \sqrt{F}} 
       \Big[ \tilde c_{1 \alpha\beta} ({\bf k}) H_{\nu_g}^{(1)} (k|\eta|)
       + \tilde c_{2 \alpha\beta} ({\bf k}) H_{\nu_g}^{(2)} (k|\eta|) \Big], 
       \quad \nu_g \equiv \sqrt{ n_g + {1 \over 4} }.
   \label{GW-exact-sol}
\eea
Using parameters
\bea
   \epsilon_1 \equiv {\dot H \over H^2}, \quad
       \epsilon_3 \equiv {1 \over 2} {\dot F \over H F}, \quad
   \label{epsilons}
\eea
we have
\bea
   {z_g^{\prime\prime} \over z_g} = a^2 H^2 \left( 1 + \epsilon_3 \right)
       \left( 2 + \epsilon_1 + \epsilon_3 \right)
       + a^2 H \dot \epsilon_3.
\eea
{}For $\dot \epsilon_i = 0$ we have
\bea
   n_g = { ( 1 + \epsilon_3 ) ( 2 + \epsilon_1 + \epsilon_3 )
       \over (1 + \epsilon_1)^2 },
   \label{n_g}
\eea
and in this case we have the exact solution in Eq. (\ref{GW-mode-sol}).
In Sec. \ref{sec:Applications} we will see that most of the currently favored 
prototype inflation models satisfy this $n_g = {\rm constant}$ condition.

We introduce a decomposition based on the two polarization states
\bea
   C_{\alpha\beta}^{(t)} ({\bf x}, t)
   &\equiv& \sqrt{\rm Vol} \int {d^3 k \over (2 \pi)^{3} }
       C_{\alpha\beta}^{(t)} ({\bf x}, t; {\bf k})
   \nonumber \\
   &\equiv& \sqrt{\rm Vol} \int {d^3 k \over (2 \pi)^{3} } 
       \sum_\ell e^{i {\bf k} \cdot {\bf x}} 
       h_{\ell {\bf k}} (t) e_{\alpha\beta}^{(\ell)} ({\bf k}), 
   \label{C-decomp-c}
\eea
where ${\ell} = +, \times$;
$e_{\alpha\beta}^{(+)}$ and $e_{\alpha\beta}^{(\times)}$ are
bases of plus ($+$) and cross ($\times$) polarization states with
\bea
   e^{(\ell)}_{\alpha\beta} ({\bf k}) e^{(\ell^\prime)\alpha\beta} ({\bf k}) 
       = 2 \delta_{\ell \ell^\prime}.
\eea
[In \cite{GGT-H} we used a transverse and tracefree harmonic function
$C_{\alpha\beta}^{(t)} \equiv H_T (t) Y_{\alpha\beta}^{(t)}$,
see \cite{Bardeen}.]
We introduce
\bea
   h_{\ell} ({\bf x}, t)
   &\equiv& {1 \over 2} \sqrt{\rm Vol} \int {d^3 k \over (2 \pi)^{3}}
       C_{\alpha\beta}^{(t)} ({\bf x}, t; {\bf k})
       e^{({\ell})\alpha\beta} ({\bf k})
   \nonumber \\
   &=& \sqrt{\rm Vol} \int {d^3 k \over (2 \pi)^{3} } 
       e^{i {\bf k} \cdot {\bf x}} h_{\ell {\bf k}} (t).
   \label{h-def-c}
\eea
The classical power spectrum is introduced as
\bea
   {\cal P}_{C^{(t)}_{\alpha\beta}} ({\bf k}, t)
   \equiv {k^3 \over 2 \pi^2} \int
       \langle C_{\alpha\beta}^{(t)} ({\bf x} + {\bf r}, t)
       C^{(t)\alpha\beta} ({\bf x}, t) \rangle_{\bf x}
       e^{-i{\bf k} \cdot {\bf r}} d^3 r,
   \label{Power-def-c}
\eea
where $\langle \rangle_{\bf x}$ is a spatial average over ${\bf x}$.
We can show that
\bea
   {\cal P}_{C^{(t)}_{\alpha\beta}} ({\bf k}, t)
       = 2 \sum_\ell {\cal P}_{h_\ell} ({\bf k}, t)
       = 2 \sum_\ell {k^3 \over 2 \pi^2} \big| h_{\ell {\bf k}} (t) \big|^2.
   \label{Power-c}
\eea
In a space without any preferred direction we have
$h_{+{\bf k}} = h_{\times{\bf k}} = h_{\bf k}$;
thus, the two polarization states contribute equally.

It is well known that the various generalized gravity theories which are
subset of Eq. (\ref{GGT-action}) can be transformed to Einstein gravity
with a minimally coupled scalar field by conformal rescaling of the metric
and field \cite{CT,GGT-1990}.
In \cite{GGT-1990,GGT-CT} we derived the background and perturbed sets of 
equations valid in the generalized gravity theories by conformal rescaling
of the known results in Einstein gravity.
{}From \cite{GGT-1990,GGT-CT} we find:
\bea
   \tilde g_{ab} \equiv \Omega^2 g_{ab}, \quad 
       \Omega \equiv \sqrt{F}, \quad
       \tilde a = \Omega a, \quad
       d \tilde t = \Omega d t, \quad 
       \tilde \eta = \eta, \quad
       \tilde C_{\alpha\beta}^{(t)} = C_{\alpha\beta}^{(t)}, \quad
       \tilde {\bf k} = {\bf k},
   \label{CT-1}
\eea
where the tilde indicates a quantity in the conformally transformed 
Einstein frame.
Using Eq. (\ref{CT-1}) we can show that the conformal transformation
of Eqs. (\ref{GW-action}-\ref{Power-c}) correctly reproduce the equations 
valid in Einstein gravity, \cite{GGT-CT}.

In the matter dominated era with Einstein's gravity we have
$F = 1/(8 \pi G)$ and $a \propto t^{2/3}$.
{}From Eq. (\ref{epsilons}) we have $\epsilon_1 = - {3 \over 2}$ 
and $\epsilon_3 = 0$, thus $n_g = 2$ and $\nu_g = {3 \over 2}$.
In terms of the spherical Bessel function we have
$h_{{\bf k}} \propto j_1(k\eta)/(k\eta), \; n_1(k\eta)/(k\eta)$.
After a long period of evolution in the superhorizon scale ($k\eta \ll 1$)
only the growing mode (the first term) will survive.
Conventionally, we let
\bea
   h_{{\bf k}} (t) \equiv A_T^{1/2} ({\bf k}) k^{-3/2} 
       \left[ {3 j_1(k \eta) \over k \eta} \right].
   \label{h_k-MDE}
\eea
{}From Eq. (\ref{Power-c}) the power spectrum becomes
\bea
   {\cal P}^{1/2}_{C^{(t)}_{\alpha\beta}} ({\bf k}, t)
       = {\sqrt{2} \over \pi} A_T^{1/2} ({\bf k}) 
       \left| {3 j_1(k \eta) \over k \eta} \right|.
   \label{P-MDE}
\eea
One often writes
\bea
   A_T ({\bf k}) \equiv A_T k^{n_T}.
   \label{n_T}
\eea
At the horizon crossing epoch (${k \over aH}|_{\rm HC} \equiv 1$)
we have $k \eta = 2 {k \over aH} = 2$, and thus a 
{\it scale invariant spectrum} corresponds to $n_T = 0$.

\section{Quantum generation}
                               \label{sec:Quantum}

The accelerated expansion stage (inflation) can
generate the stochastic gravitational wave by rapidly stretching the quantum
vacuum fluctuations of the perturbed metric to superhorizon scale.
In order to handle the quantum mechanical generation of the
gravitational wave we consider Hilbert space operator 
$\hat C^{(t)}_{\alpha\beta}$ instead of the classical metric perturbation
$C^{(t)}_{\alpha\beta}$.
We write (see \cite{Allen})
\bea
   \hat C_{\alpha\beta}^{(t)} ({\bf x}, t)
   &\equiv& \int {d^3 k \over (2 \pi)^{3/2} }
       \hat C_{\alpha\beta}^{(t)} ({\bf x}, t; {\bf k})
   \nonumber \\
   &\equiv& \int {d^3 k \over (2 \pi)^{3/2} } 
       \left[ \sum_\ell e^{i {\bf k} \cdot {\bf x}} \tilde h_{\ell {\bf k}} (t) 
       \hat a_{\ell {\bf k}} e_{\alpha\beta}^{(\ell)} ({\bf k}) 
       + {\rm h.c.} \right].
   \label{C-decomp-q}
\eea
In order to distinguish the mode function $\tilde h_{\ell {\bf k}} (t)$
from the classical one in Eq. (\ref{C-decomp-c}) we have put a tilde on it.
The creation and annihilation operators of each polarization state follow
\bea
   [ \hat a_{\ell {\bf k}}, \hat a^\dagger_{\ell^\prime {\bf k}^\prime} ] 
       = \delta_{\ell \ell^\prime} \delta^3 ({\bf k} - {\bf k}^\prime),
\eea
and zero otherwise.
By introducing
\bea
   \hat h_{\ell} ({\bf x}, t)
   &\equiv& {1 \over 2} \int {d^3 k \over (2 \pi)^{3/2}}
       \hat C_{\alpha\beta}^{(t)} ({\bf x}, t; {\bf k})
       e^{({\ell})\alpha\beta} ({\bf k})
   \nonumber \\
   &=& \int {d^3 k \over (2 \pi)^{3/2} } \Big[ e^{i {\bf k} \cdot {\bf x}}
       \tilde h_{\ell {\bf k}} (t) \hat a_{\ell {\bf k}} + {\rm h.c.} \Big],
   \label{h-def-q}
\eea
Eq. (\ref{GW-action}) can be written as
\bea
   \delta^2 S_g 
       = \int a^3 F \sum_{\ell} \left( \dot {\hat h}_{\ell}^2
       - {1 \over a^2} \hat h_{\ell}^{\;\; |\gamma} 
       \hat h_{{\ell},\gamma} \right) dt d^3 x.
   \label{GW-action-h} 
\eea
The equation of motion becomes
\bea
   \ddot {\hat h}_\ell 
       + \left( 3 H + {\dot F \over F} \right) \dot {\hat h}_\ell
       - {1 \over a^2} \nabla^2 \hat h_\ell = 0.
   \label{GW-eq}
\eea
The conjugate momenta are
\bea
   \delta \hat \pi_{h_{\ell}} ({\bf x}, t) 
       = {\partial {\cal L} \over \partial \dot {\hat h}_{\ell}}
       = 2 a^3 F \dot {\hat h}_{\ell}.
\eea
{}From the equal time commutation relation between
$\hat h_{\ell}$ and $\delta \hat \pi_{h_{\ell}}$ we have
\bea
   [ \hat h_{\ell} ({\bf x}, t), \dot {\hat h}_{\ell} ({\bf x}^\prime, t) ] 
       = {i \over 2 a^3 F} \delta^3 ( {\bf x} - {\bf x}^\prime ).
   \label{commutation}
\eea
{}From Eqs. (\ref{h-def-q},\ref{commutation}) we have
\bea
   \tilde h_{\ell {\bf k}} (t) \dot {\tilde h}^*_{\ell {\bf k}} (t) 
       - {\tilde h}^*_{\ell {\bf k}} (t) \dot {\tilde h}_{\ell {\bf k}} (t)
       = {i \over 2 a^3 F}.
   \label{h-normalization}
\eea

{}For $n_g = {\rm constant}$ Eq. (\ref{GW-eq}) has an exact solution
as in Eq. (\ref{GW-exact-sol}).
In terms of the mode function we have
\bea
   \tilde h_{\ell {\bf k}} (\eta) = {\sqrt{ \pi |\eta|} \over 2 a} 
       \Big[ c_{\ell 1} ({\bf k}) H_{\nu_g}^{(1)} (k|\eta|)
       + c_{\ell 2} ({\bf k}) H_{\nu_g}^{(2)} (k|\eta|) \Big] 
       \sqrt{ 1 \over 2 F}, 
   \label{GW-mode-sol}
\eea
where according to the normalization condition in Eq. (\ref{h-normalization})
the coefficients $c_{\ell 1} ({\bf k})$ and $c_{\ell 2} ({\bf k})$ follow 
\bea
   \left| c_{\ell 2} ({\bf k}) \right|^2 
       - \left| c_{\ell 1} ({\bf k}) \right|^2 = 1.
   \label{c-normalization}
\eea
We introduce the power spectrum of the Hilbert space graviational wave 
operator based on the vacuum expectation value as
\bea
   {\cal P}_{\hat C^{(t)}_{\alpha\beta}} ({\bf k}, t)
   \equiv {k^3 \over 2 \pi^2} \int
       \langle \hat C_{\alpha\beta}^{(t)} ({\bf x} + {\bf r}, t)
       \hat C^{(t)\alpha\beta} ({\bf x}, t) \rangle_{\rm vac} 
       e^{-i{\bf k} \cdot {\bf r}} d^3 r,
   \label{Power-def-q}
\eea
where $\langle \rangle_{\rm vac} \equiv \langle {\rm vac}| | {\rm vac} \rangle$
is a vacuum expectation value with
$\hat a_{\ell {\bf k}} |{\rm vac} \rangle \equiv 0$ for all ${\bf k}$.
We can show
\bea
   {\cal P}_{\hat C^{(t)}_{\alpha\beta}} ({\bf k}, t)
       = 2 \sum_\ell {\cal P}_{\hat h_\ell} ({\bf k}, t)
       = 2 \sum_\ell {k^3 \over 2 \pi^2} 
       \big| \tilde h_{\ell {\bf k}} (t) \big|^2.
   \label{Power-q}
\eea

In the Einstein gravity limit, where $F = 1/(8 \pi G)$, each $\hat h_\ell$ in 
Eq. (\ref{GW-action-h}) can be corresponded to a minimally coupled scalar 
field without potential with a normalization 
$\hat h_\ell = \hat \phi /\sqrt{2F} = \sqrt{4 \pi G} \hat \phi$;
in this case, assuming equal contributions from each polarization,
the power spectrum becomes 
\bea
   {\cal P}^{1/2}_{\hat C^{(t)}_{\alpha\beta}} 
       = 2 {\cal P}^{1/2}_{\hat h_\ell} 
       = \sqrt{16 \pi G} {\cal P}^{1/2}_{\hat \phi}.
   \label{Power-MSF}
\eea

Using Eq. (\ref{CT-1}) we can show that the conformal transformations
of Eqs. (\ref{C-decomp-q}-\ref{Power-q}) also reproduce the equations 
valid in Einstein gravity.

\section{Vacuum Fluctuations}
                                    \label{sec:Applications}

In this section we will apply the quantum formulation of the graviational 
wave in generalized gravity to various specific cosmological situations.
We will derive the quantum vacuum fluctuations in several interesting
cosmological models in analytic forms.
This analytic treatments are possible mainly because the background models
will satisfy the $n_g = {\rm constant}$ condition, thus allowing
exact solutions in Eq. (\ref{GW-mode-sol}).
The corresponding vacuum fluctuations of the scalar type perturbation
are presented in \cite{Kin}.
Since the equations and the solutions of the background models are presented 
in Sec. III of \cite{Kin}, in the following we summarize the background 
solutions without deriving them again.

\subsection{Minimally coupled scalar field}
                                   \label{sec:MSF}

The minimally coupled scalar field is a case of Eq. (\ref{GGT-action}) with
$f = R/(8 \pi G)$, thus $F = 1/(8 \pi G)$, and $\omega = 1$.
We have $\epsilon_3 = 0$.
The cases where the background scale factor follows an exponential or 
a power-law expansion in time correspond to $\dot \epsilon_1 = 0$.
Thus $n_g$ is a constant and the solution in Eq. (\ref{GW-mode-sol}) applies.

\subsubsection{Exponential expansion} 

{}For $a \propto e^{Ht}$ with $H = {\rm constant}$ the background has 
a solution with $V = {\rm constant}$ and $\dot \phi = 0$.
We have $\epsilon_1 = 0$, thus $n_g = 2$ and $\nu_g = {3 \over 2}$.
{}From Eqs. (\ref{GW-mode-sol},\ref{Power-q}) we can present the power spectrum 
valid in general scale.
In the large scale limit we have
\bea
   {\cal P}^{1/2}_{\hat C^{(t)}_{\alpha\beta}} ({\bf k}, \eta)
       = \sqrt{16 \pi G} {H \over 2 \pi}
       \sqrt{ {1 \over 2} \sum_\ell \Big| c_{\ell 2} ({\bf k})
       - c_{\ell 1} ({\bf k}) \Big|^2 }.
   \label{P-EXP}
\eea
 
In Eq. (\ref{P-EXP}) and in the following, $c_{\ell 1} ({\bf k})$
and $c_{\ell 2} ({\bf k})$ satisfy Eq. (\ref{c-normalization}).
{\it If} we have equal contributions from the two polarization states,
the dependence on the vacuum state in Eq. (\ref{P-EXP}) and the following
can be written as
\bea
   \sqrt{ {1 \over 2} \sum_\ell
       \Big| c_{\ell 2} ({\bf k}) - c_{\ell 1} ({\bf k}) \Big|^2 }
       = \Big| c_{\ell 2} ({\bf k}) - c_{\ell 1} ({\bf k}) \Big|.
\eea
The often favored vacuum state in the literature corresponds to a special
(and the simplest) case with $c_{\ell 2}({\bf k}) \equiv 1$ 
and $c_{\ell 1}({\bf k}) \equiv 0$, thus
\bea
   \sqrt{ {1 \over 2} \sum_\ell
       \Big| c_{\ell 2} ({\bf k}) - c_{\ell 1} ({\bf k}) \Big|^2 } = 1.
\eea
Equation (\ref{P-EXP}) with the simplest vacuum choice was derived in
\cite{pre-COBE,Abbott-Wise,Starobinsky,Bogoliubov,post-COBE,Allen}.

\subsubsection{Power-law expansion} 

{}For $a \propto t^p$ with $p = {\rm constant} > 1$ the background
has a solution with $\dot \phi = \sqrt{2p}/t$ and 
$V = p (3p -1)/t^2 \propto e^{-\sqrt{2/p} \phi}$, \cite{POW-infl}.
In this case we have $\epsilon_1 = - 1/p$, 
thus $\nu_g = {3 p - 1 \over 2(p - 1)}$;
$\nu_g$ is the same as the corresponding index appearing in the scalar
type perturbation in \cite{Kin}.
The general power spectrum follows from Eqs. (\ref{GW-mode-sol},\ref{Power-q}). 
In the large scale limit we have
\bea
   {\cal P}^{1/2}_{\hat C^{(t)}_{\alpha\beta}} ({\bf k}, \eta)
   = \sqrt{16 \pi G} 
       {H \over 2 \pi} { \Gamma(\nu_g) \over \Gamma(3/2) } { p - 1 \over p }
       \left( {2 \over k |\eta|} \right)^{\nu_g - 3/2}
       \sqrt{ {1 \over 2} \sum_\ell \Big| c_{\ell 2} ({\bf k}) 
       - c_{\ell 1} ({\bf k}) \Big|^2 }.
   \label{P-POW} 
\eea
In the limit of $p \rightarrow \infty$ Eq. (\ref{P-POW})
reproduces Eq. (\ref{P-EXP}). 
Equation (\ref{P-POW}) in the simplest vacuum choice was first derived in 
\cite{Abbott-Wise}.

\subsubsection{Potential-less case}
                                     \label{sec:Potential-less-MSF}

In a case with vanishing potential, $V(\phi) = 0$, we have a solution
with $a \propto t^{1/3}$ which corresponds to a stiffest equation of state
in the case of ideal fluid with $p = \mu$.
This case does not produce any inflation.
However, we could calculate the exact power spectrum based on the vacuum
expectation value.
We have $\epsilon_1 = -3$, thus $n_g = - {1 \over 4}$ and $\nu_g = 0$.
The general power spectrum follows from Eqs. (\ref{GW-mode-sol},\ref{Power-q}). 
In the large scale limit we have
\bea
   {\cal P}^{1/2}_{\hat C^{(t)}_{\alpha\beta}} ({\bf k}, \eta)
   = \sqrt{16 \pi G} 
       \left( { \sqrt{4 |\eta| \over a^2 } } \right)_1
       \left( {k \over 2 \pi} \right)^{3/2} \ln{(k|\eta|)} 
       \sqrt{ {1 \over 2} \sum_\ell \Big| c_{\ell 2} ({\bf k}) 
       - c_{\ell 1} ({\bf k}) \Big|^2 },
   \label{P-MSF-Kin}
\eea
where the subindex $1$ indicates that since the quantity does not depend 
on time we have evaluated it in an arbitrary time $t_1$.
Later in this section we will see that the potential-less situations
in several generalized gravity theories also produce the same
$k^3$ dependence of the power spectrums as in Eq. (\ref{P-MSF-Kin}).

\subsection{Scalar-tensor theory}
                                  \label{sec:ST}

A scalar-tensor theory is given by an action \cite{BD}
\bea
   S = \int d^4 x \sqrt{-g} \left[ \phi R
       - \omega (\phi) { \phi^{;a} \phi_{,a} \over \phi} - V(\phi) \right],
   \label{ST-action}
\eea
which is a case of Eq. (\ref{GGT-action}) with $F = 2 \phi$.
{\it Ignoring} the potential term and for $\omega = {\rm constant}$ we have
\cite{Levin}:
\bea
   & & a \propto | t_0 - t |^{ - q }, \quad
       \phi \propto | t_0 - t |^{ 1 + 3q }, \quad
       q \equiv - { 1 + \omega \mp \sqrt{1 + {2 \over 3} \omega}
       \over 4 + 3 \omega }.
   \label{BG-sol-ST}
\eea
A pole-like acceleration stage can be realized when $q>0$ which corresponds 
to the upper sign and $t_0 > t$.
In the following analyses, for generality we will consider {\it both} signs. 
{}From Eq. (\ref{epsilons}) we have
$\epsilon_1 = 1/q$ and $\epsilon_3 = - (1 + 3 q)/(2q)$,
thus $n_g = - {1 \over 4}$ and $\nu_g = 0$.
The general power spectrum follows from Eqs. (\ref{GW-mode-sol},\ref{Power-q}). 
In the large scale limit we have
\bea
   {\cal P}^{1/2}_{ \hat C_{\alpha\beta}^{(t)} } ({\bf k}, \eta)
   = \left( \sqrt{ 4 |\eta| \over a^2 \phi } \right)_1
       \left( {k \over 2 \pi} \right)^{3/2} \ln{(k|\eta|)} 
       \sqrt{ {1 \over 2} \sum_\ell \Big| c_{\ell 2} ({\bf k}) 
       - c_{\ell 1} ({\bf k}) \Big|^2 }.
   \label{P-ST} 
\eea

\subsection{Induced gravity theory}
                                  \label{sec:Induced}

The induced gravity theory is given by \cite{Zee}
\bea
   S = \int d^4 x \sqrt{-g} \left[ {1 \over 2} \epsilon \phi^2 R
       - {1 \over 2} \phi^{;a} \phi_{,a} - V(\phi) \right],
   \label{Induced-action}
\eea
which is a case of Eq. (\ref{GGT-action}) with $F = \epsilon \phi^2$.
{\it Assuming} $V = 0$ we have the following background solution 
\cite{Pollock}:
\bea
   & & a \propto | t_0 - t |^{-q}, \quad
       \phi \propto | t_0 - t |^{1 + 3q \over 2}, \quad
       q \equiv - { 1 + 4 \epsilon \mp 4 \epsilon 
       \sqrt{ 1 + {1 \over 6 \epsilon} } \over 3 + 16 \epsilon}.
   \label{BG-sol-Induced}
\eea
Cases with $q>0$, thus the upper sign, and $t_0 > t$ include the pole-like 
acceleration stage.
{}For generality, we will take both signs.
{}From Eq. (\ref{epsilons}) we have $\epsilon_1 = 1/q$ and 
$\epsilon_3 = - (1 + 3q)/(2q)$, thus $n_g = - {1 \over 4}$ and $\nu_g = 0$.
The general power spectrum follows from Eqs. (\ref{GW-mode-sol},\ref{Power-q}). 
In the large scale limit we have
\bea
   {\cal P}^{1/2}_{ \hat C_{\alpha\beta}^{(t)} } ({\bf k}, \eta)
   = \left( { \sqrt{ 8 |\eta| \over \epsilon a^2 \phi^2 } } \right)_1
       \left( {k \over 2 \pi} \right)^{3/2} \ln{(k|\eta|)} 
       \sqrt{ {1 \over 2} \sum_\ell \Big| c_{\ell 2} ({\bf k}) 
       - c_{\ell 1} ({\bf k}) \Big|^2 }.
   \label{P-Induced}
\eea

\subsection{String theory}
                                  \label{sec:String}

The low-energy effective action of string theory is given by
\cite{string-action}
\bea
   S = \int d^4 x \sqrt{-g} {1 \over 2} e^{-\phi}
       \left( R + \phi^{;a} \phi_{,a} \right),
   \label{String-action}
\eea
which is a case of Eq. (\ref{GGT-action}) with $F = e^{-\phi}$.
{}For the background we have a solution
\bea
   a \propto | t_0 - t |^{\mp 1/\sqrt{3} }, \quad
       e^\phi \propto | t_0 - t |^{-1 \mp \sqrt{3} }.
   \label{BG-sol-string}
\eea
The upper sign with $t < t_0$ represents a pole-like inflation stage
which is called as a pre-big bang stage in \cite{pre-big-bang}.
In the following we consider both signs for generality.
{}From Eq. (\ref{epsilons}) we have
$\epsilon_1 = \pm \sqrt{3}$ and $\epsilon_3 = - ( 3 \pm \sqrt{3} )/2$,
thus $n_g = - {1 \over 4}$ and $\nu_g = 0$.
The general power spectrum follows from Eqs. (\ref{GW-mode-sol},\ref{Power-q}). 
In the large scale limit we have
\bea
   {\cal P}^{1/2}_{ \hat C_{\alpha\beta}^{(t)} } ({\bf k}, \eta)
   = \left( \sqrt{ 8 |\eta| \over a^2 e^{-\phi} } \right)_1
       \left( {k \over 2 \pi} \right)^{3/2} \ln{(k|\eta|)} 
       \sqrt{ {1 \over 2} \sum_\ell \Big| c_{\ell 2} ({\bf k}) 
       - c_{\ell 1} ({\bf k}) \Big|^2 }.
   \label{P-string} 
\eea
The gravitational wave power spectrum in the pre-big bang scenario has been 
studied in \cite{pre-big-bang-GW}.

\subsection{Conformal transformation}
                                  \label{sec:CT}

As generally have shown in Secs. \ref{sec:Classical},\ref{sec:Quantum}
the conformal transformation relates the results between the
generalized gravity theories and Einstein gravity.
{}For individual generalized gravity theories in 
Secs. \ref{sec:ST}-\ref{sec:String}, using the conformal transformation
properties in Eq. (\ref{CT-1}), we can show that the conformal transformations
of the results in Secs. \ref{sec:ST}-\ref{sec:String} reproduce the
results in Sec. \ref{sec:Potential-less-MSF}.

\section{Gravitational wave spectrums}
                                      \label{sec:Spectrums}

{}From Eq. (\ref{GW-LS-sol}) we find that in the large scale limit 
the growing mode of $C^{(t)}_{\alpha\beta}$ (and $h_\ell$) is conserved
\bea
   h_\ell ({\bf x}, t) = C_\ell ({\bf x}).
   \label{h-conserv}
\eea
We emphasize that Eq. (\ref{h-conserv}) is valid for general changes in 
$V(\phi)$, $\omega (\phi)$ and $f(\phi,R)$ in generalized gravity \cite{GGT-H}, 
and the change in the equation of state $p = p(\mu)$ in the fluid era 
\cite{PRW}.
This reflects the kinematic nature of the evolution in the superhorizon scale.
Equation (\ref{h-conserv}) is valid in a scale larger than the visual horizon.

The gravitational waves in the observationally relevant scales are
stretched into superhorizon scale during the inflation era.
The evolution of the gravitational wave in the superhorizon scale  
is characterized by the above temporal conservation.
Thus, the classical spectrum of the gravitational wave in later epochs
can be easily derived.
In the literature, the popular method is based on matching the Bogoliuvov 
coefficients assuming the sudden jump transitions among different cosmological 
eras (the inflation era, the radiation era, and the matter era) 
\cite{Bogoliubov,Allen}; in the case of a scalar type perturbation
the sudden jump approximation using the joining variables is made in \cite{HV}.
However, since the conservation variable is applicable considering
general changes in the background state of the universe (the equation of state, 
the field potential, and even the gravity theories) we can derive
the same result in a more direct manner:
As the quantum fluctuation of $\hat C_{\alpha\beta}^{(t)}$ becomes 
superhorizon scale and classicalizes, it can somehow be related to the 
classical fluctuation of $C_{\alpha\beta}^{(t)}$, and thereafter, 
simply is conserved as long as the scale remains in the superhorizon scale. 
As the fluctuation transits back into horizon scale in the matter era
in Einstein gravity, the solution in Eq. (\ref{h_k-MDE}) will apply.

Comparing Eqs. (\ref{Power-def-q},\ref{Power-q}) and
Eqs. (\ref{Power-def-c},\ref{Power-c}) we are tempted to take 
the following {\it ansatz}: 
in the large scale limit during inflation era we assume
\bea
   {\cal P}_{h_\ell} ({\bf k},\eta)
       \equiv {\cal P}_{\hat h_\ell} ({\bf k}, \eta) \times Q_\ell ({\bf k}),
   \label{ansatz}
\eea
where $Q_\ell ({\bf k})$ is a classicalization factor for the 
gravitational wave with a polarization state $\ell$.
$Q_\ell ({\bf k})$ may take into account of the possible nontrivial 
effects which arise during the classicalization processes of the 
fluctuating quantum gravitational wave field, \cite{Hu}.
If we assume equal contributions from both polarizations, we have
\bea
   {\cal P}^{1/2}_{C^{(t)}_{\alpha\beta}} ({\bf k}, \eta)
       = {\cal P}^{1/2}_{\hat C^{(t)}_{\alpha\beta}} ({\bf k}, \eta)
       \times \sqrt{Q_\ell (k)}.
   \label{ansatz-2}
\eea
In Eqs. (\ref{ansatz},\ref{ansatz-2}) the right hand side should be evaluated
while the scale is in the large scale limit during the inflation era.
The power spectrums of the vacuum fluctuations, 
${\cal P}_{\hat C^{(t)}_{\alpha\beta}}$, 
for several expansion stages are derived in Sec. \ref{sec:Applications}.

The classicalized gravitational wave in the inflation era 
is, later on, preserved as in Eq. (\ref{h-conserv}) as long as 
the scale remains in the superhorizon scale.
As emphasized, the conserved behavior holds including the transitions
between the inflation era (possibly based on generalized gravity) 
and the radiation era, and between the radiation era and the matter era.
During the matter era we have a solution in Eq. (\ref{h_k-MDE}).
Thus, by directly comparing Eqs. (\ref{P-MDE},\ref{ansatz})
in the superhorizon scales, and using one of the power spectrums 
in Sec. \ref{sec:Applications} as the seed generating mechanism,
we can determine the classical amplitude of the gravitational wave spectrum, 
$A_T ({\bf k})$, in Eq. (\ref{n_T}).
As an example, in a scenario with exponential inflation based on 
Einstein gravity, from Eqs. (\ref{P-MDE},\ref{ansatz},\ref{P-EXP})
we can show that 
\bea
   A_T^{1/2} ({\bf k})
   = {\pi \over \sqrt{2}} \sqrt{16 \pi G} {H \over 2 \pi}
       \sqrt{ {1 \over 2} \sum_\ell \Big| c_{\ell 2} ({\bf k})
       - c_{\ell 1} ({\bf k}) \Big|^2 \times Q_\ell ({\bf k}) }.
   \label{P-EXP-A_T}
\eea
If we ignore the dependences on the choice of the vacuum state and the 
classicalization factor, we have 
\bea
   \sqrt{ {1 \over 2} \sum_\ell \Big| c_{\ell 2} ({\bf k})
       - c_{\ell 1} ({\bf k}) \Big|^2 \times Q_\ell ({\bf k}) } = 1.
\eea
In this case we have $A_T({\bf k}) = 2 \pi G H^2 \propto k^0$.
Thus, from Eq. (\ref{n_T}) we have a scale invariant spectrum with $n_T = 0$.
Similarly, in the power-law inflation in Einstein gravity, using
Eq. (\ref{P-POW}) we have $n_T = 3 - 2 \nu_g = 2 /(1 - p)$.
In the potential-less case in Eq. (\ref{P-MSF-Kin})
and also in the various pole-like inflation models 
in Eqs. (\ref{P-ST},\ref{P-Induced},\ref{P-string}), 
additionally ignoring the logarithmically dependent term,
we have a common spectrum with
\bea
   n_T = 3.
\eea
This spectrum fundamentally differs from the scale invariant one 
with $n_T = 0$.
We, however, point out that the choice of the physically relevant vacuum
state and the classicalization factor could possibly dominate both the 
amplitude and the spectral dependence of the generated power spectrum.

The propagation of the relic radiation will be affected by the classically 
fluctuating {\it spacetime} caused by the classical gravitational wave.
In our scenario the classical wave may have the origin in the inflation era
as the linear order quantum fluctuations of the spacetime
are stretched to superhorizon scale and are classicalized.
The effects of the curved spacetime (including effects due to the 
graviational wave) on the propagation of the relic cosmic microwave 
background photons cause an important observational consequence.
The gravitational redshifts of the relic photons cause temperature fluctuations
of the present day microwave photons in different directions \cite{Sachs-Wolfe}
\bea
   {\delta T \over T} ({\bf e}; {\bf x}_R) 
       = - \int_{\lambda_e}^{\lambda_o} \left[ {\partial \over \partial \eta}
       C^{(t)}_{\alpha\beta} ({\bf x}(\lambda), \eta) \right]_{\eta = \lambda}
       e^\alpha e^\beta d \lambda,
   \label{Sachs}
\eea
where ${\bf e}$ is a unit vector in the direction of the relic photon;
$\lambda_e = \lambda_{\rm emitted}$ and $\lambda_o = \lambda_{\rm observed}$.
We have parametrized the path of the photon so that 
${\bf x} (\lambda) = x(\lambda) {\bf e} - {\bf x}_R$, 
where $x (\lambda) = \eta_o - \lambda$ and ${\bf x}_R$ is the location
of the observer.
We expand
\bea 
   {\delta T \over T} ({\bf e}; {\bf x}_R) 
       = \sum_{lm} a_{lm} ({\bf x}_R) Y_{lm} ({\bf e}).  
   \label{a_lm} 
\eea 
We can derive the rotationally symmetric quantity $\langle a_l^2 \rangle$ 
$\equiv$ $\langle |a_{lm} ({\bf x}_R)|^2 \rangle_{{\bf x}_R}$,
where $\langle \rangle_{{\bf x}_R}$ is an average over every possible location
of the observer.
Using Eqs. (\ref{a_lm},\ref{Sachs},\ref{C-decomp-c}), and after a little 
but well known algebra (for example, see \cite{Allen}), we can derive
\cite{Abbott-Wise,Starobinsky}
\bea
   \langle a_l^2 \rangle
       &=& {2 \over \pi} {\Gamma(l + 3) \over \Gamma(l -1)}
       \int_0^\infty \Bigg| 
       \int_{\eta_e}^{\eta_o}
       {\partial h_{\bf k} (\eta) \over \partial \eta}
       {j_l (k \eta_0 - k \eta) \over (k \eta_0 - k \eta )^2} k d \eta
       \Bigg|^2 d k
   \nonumber \\
       &=& {9 \pi \over 2} {\Gamma(l + 3) \over \Gamma(l -1)}
       \int_0^\infty A_T (k) |I_l (k)|^2 d \ln{k}, 
   \nonumber \\
   & & I_l (k) \equiv {2 \over \pi} \int_{\eta_e}^{\eta_o}
       {j_2 (k \eta) \over k \eta}
       {j_l (k \eta_0 - k \eta) \over (k \eta_0 - k \eta )^2} k d \eta,
   \label{a_l}
\eea
where in the second step we used Eq. (\ref{h_k-MDE}).
In Eq. (\ref{a_l}) we have assumed equal contributions from both polarizations.
In combination with the derived $A_T({\bf k})$ for each seed generating
mechanism in Sec. \ref{sec:Applications}, for example Eq. (\ref{P-EXP-A_T})
in the exponential inflation scenario, the multipole moments of
the temperature anisotropy can be known by evaluating Eq. (\ref{a_l}).

\section{Discussions}
                           \label{sec:Discussions}

In this paper we have considered the inflation models based on 
generalized gravity theories.
We have derived the seed spectrums of the graviational wave based on the
vacuum expectations of the quantized fluctuating metric in analytic forms.
In generalized gravity theories the general background evolutions 
under the potential-less assumption lead to common spectrums for
the gravitational wave and for the scalar type perturbation as
\bea
   n_T \simeq 3, \quad n \simeq 4,
\eea
whereas the observationally favored \cite{COBE} scale-invariant spectrums 
based on the near exponential inflation are $n_T \simeq 0$ and $n \simeq 1$;
$n$ is a spectral index for the density perturbation defined in \cite{Kin}.

The proper [which means the mathematical convenience at least] 
handling of the scalar type perturbation is possible by a following
gauge invariant combination of variables
\bea
   \delta \phi_\varphi \equiv \delta \phi - {\dot \phi \over H} \varphi
       \equiv - {\dot \phi \over H} \varphi_{\delta \phi},
\eea
where $\delta \phi ({\bf x}, t)$ is the perturbed part of the scalar 
(or dilaton) field, and $\varphi ({\bf x}, t)$ is the perturbed part of 
the spatial scalar curvature, see Eqs. (3,4) of \cite{PRW}.
Thus the gauge invariant combinations $\delta \phi_\varphi$ and 
$\varphi_{\delta \phi}$ are the same as $\delta \phi$ in the uniform-curvature 
gauge and $\varphi$ in the uniform-field gauge, respectively;
the uniform-field gauge coincides with the comoving gauge in the limit
of the minimally coupled scalar field.
As thoroughly presented in \cite{GGT-H,GGT-QFT,Kin} the scalar type 
perturbation in the large scale limit is also characterized by a conserved 
quantity, which is $\varphi_{\delta \phi}$.
We also have general asymptotic solutions and the exact analytic solution 
under a certain, but generally applicable, condition.
By comparing the power spectrums in Sec. \ref{sec:Applications} with the ones
of the scalar type perturbation in Sec. III of \cite{Kin} we can show the 
following strikingly similar structures.
In the three cases of the minimally coupled scalar field in 
Sec. \ref{sec:MSF}, by taking the simplest vacuum choice, we have
\bea
   {\cal P}^{1/2}_{\hat C_{\alpha\beta}^{(t)}} ({\bf k}, t)
       = \sqrt{16 \pi G} \times
       {\cal P}^{1/2}_{\delta \hat \phi_{\varphi}} ({\bf k}, t).
\eea
Similarly, in the three cases of the generalized gravity theories in 
Secs. \ref{sec:ST}-\ref{sec:String} and the case in 
Sec. \ref{sec:Potential-less-MSF}, also by taking the simplest vacuum choices, 
we have
\bea
   {\cal P}^{1/2}_{\hat C_{\alpha\beta}^{(t)}} ({\bf k}, t)
       = 2 \sqrt{3} \times
       {\cal P}^{1/2}_{\hat \varphi_{\delta \phi}} ({\bf k}, t).
\eea

Recently, there have been many attempts to reconstruct the inflationary 
potential from the observed large scale structure and the microwave anisotropy,
\cite{Reconstruct}.
These attempts aim a high accuracy reconstruction of the inflaton potential. 
In such cases one may be not allowed to ignore two potentially important 
effects which have been always ignored in the literature: the spectral 
dependences on the vacuum choice and on the classicalization factor.
These two effects could separately occur for both the gravitational wave
and the scalar type perturbation \cite{Kin}, and could possibly
dominate the spectrums.
The accurate determination of the classicalization factors may require the 
treatment which goes beyond the linear approximation of the quantum 
fluctuations.
The vacuum choices in the curved spacetime may also depend on
concrete physical arguments.

We would like to emphasize that the formulation made in 
Secs. \ref{sec:Classical},\ref{sec:Quantum} in a unified way is generally 
applicable to a broad class of gravity theories included in 
Eq. (\ref{GGT-action}).

\section*{Acknowledgments}

We thank H. Noh for useful discussions.
This work was supported by the KOSEF, Grant No. 95-0702-04-01-3
and through the SRC program of SNU-CTP.


\end{document}